\documentclass[10pt,conference]{IEEEtran}
\usepackage{latexsym}
\usepackage{ngerman}
\renewcommand{\figurename}{Figure}
\renewcommand{\tablename}{Table}
\renewcommand{\abstractname}{Abstract}
\usepackage{graphicx}
\usepackage{array}
\usepackage{multirow}
\usepackage{hhline}		
\usepackage{subfigure}
\usepackage{mathptmx}
\usepackage{mathtools}
\usepackage{amsfonts}
\usepackage{amssymb}
\usepackage{amsbsy}
\usepackage{adjustbox}
\usepackage{todonotes}
\usepackage{placeins}
\ifCLASSINFOpdf
\else
\fi
\usepackage{url}

\usepackage[pscoord]{eso-pic}
        \newcommand{\placetextbox}[3]{
               \setbox0=\hbox{#3}
               \AddToShipoutPictureFG*{
               \put(\LenToUnit{#1\paperwidth},\LenToUnit{#2\paperheight}){
                       \vtop{{\null}\makebox[0pt][c]{#3}}}
               }
        }
        \placetextbox{.33}{0.06}{978-3-901882-77-7 ~\copyright~IFIP, 2015. This is the author's version of the work.}
        \placetextbox{.358}{0.048}{It is posted here by permission of IFIP for your personal use. Not for redistribution.}
        \placetextbox{.37}{0.036}{The definitive version was published in Proceedings of the International Conference on }
        \placetextbox{.419}{0.024}{Network and Service Management, 173-177, http://dl.ifip.org/db/conf/cnsm/cnsm2015/1570163515.pdf}
        
\begin{document}
\title{Modelling of IP Geolocation by use of Latency Measurements}
\author{
\IEEEauthorblockN{Peter Hillmann, Lars Stiemert, Gabi Dreo Rodosek, and Oliver Rose}
\IEEEauthorblockA{Research Center CODE (Cyber Defence)\\
	Universit\"at der Bundeswehr M\"unchen\\
Neubiberg, 85577, GERMANY\\
Email: \{peter.hillmann, lars.stiemert, gabi.dreo, oliver.rose\}@unibw.de
}
}

\maketitle

\begin{abstract}
IP Geolocation is a key enabler for many areas of application like Content Delivery Networks, targeted advertisement and law enforcement.
Therefore, an increased accuracy is needed to improve service quality.
Although IP Geolocation is an ongoing field of research for over one decade, it is still a challenging task, whereas good results are only achieved by the use of active latency measurements.
This paper presents an novel approach to find optimized Landmarks positions which are used for active probing and introduce an improved location estimation.
Since a reasonable Landmark selection is important for a highly accurate localization service, the goal is to find Landmarks close to the target with respect to the infrastructure and hop count.
Current techniques provide less information about solving this problem as well as are using imprecise models.
We demonstrate the usability of our approach in a real-world environment.
The combination of an optimized Landmark selection and advanced modulation results in an improved accuracy of IP Geolocation.
\end{abstract}


\IEEEpeerreviewmaketitle

\section{Introduction}\label{sec:introduction}
\begin{figure}[bhpt]
	\vspace*{0.2cm}
	\centering
	\includegraphics[width=0.47 \textwidth]{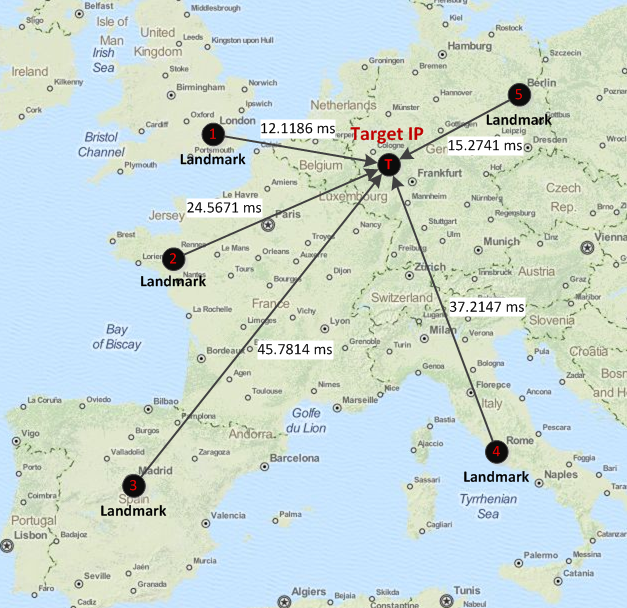}
	\caption{Example of measurement based IP localization.}
	\label{fig:GeneralMeasurement}
	\vspace*{-0.4cm}
\end{figure}

Determining the real-world geographical location of a network entity is called \textit{Geolocation}. It describes the process of allocating a physical location, e.g. defined by country, city, longitude, and latitude, to a logical address by using the Internet Protocol (IP) \cite{Padmanabhan2001}. 
The necessity for a highly accurate and reliable Geolocation service has been identified as an important goal for the Internet \cite{Katz-Bassett2006}.
More and more applications are taking into account from where users are accessing. Thereby location-aware services offer novel functionalities to their customers and provide adjusted content. One major field of application is location based advertising 
\cite{Wong2007}. Customers are automatically redirected to the appropriate language or receive, e.g. advertisement from shops in their surroundings.
Another important use case are Content Delivery Networks (CDN) \cite{Dabek2004}. In this context location information are supporting optimized load balancing between customer and mirror servers and providing better traffic management for downloads \cite{Bindal2006,Aggarwal2007}. 
As illustrated in Figure \ref{fig:GeneralMeasurement} measurement based techniques are relying on actively probing a particular host and infer the geographical location by measuring latencies.
For this purpose most of these procedures are utilizing reference hosts with well-known location information, called \textit{Landmarks} or \textit{Vantage Points}. Since these approaches are using the moderate correlation between network delay and geographic distance, the accuracy is mainly influenced by the selection of Landmarks and the mathematical modelling \cite{Ziviani2005503}.

The paper introduce two novelties. First, we present Dragoon an improvement in terms of positioning and selection of Landmarks for Geolocation strategies based on latency measurements.  
Second, we describe an advanced approach of a more accurate mathematical modelling of the location estimation process.
Therefor, the correlation between network delay, latency measurement, network topology and geographical distance is analysed for the first time focusing Europe.


\section{Related Work}\label{sec:sota}
According to Endo et al. \cite{Endo2010} approaches for IP Geolocation can be classified in either IP mapping based - including semantic - or measurement based strategies. 
In this work, we focus on the more dynamic and actual measurement based techniques. These
are relying on an active interaction with the target system.
The similarity of all those strategies is, that they are based on the assumption of an existent correlation between network latency and geographical distance.
This relationship has been proven \cite{Ziviani2005503}. 
%
\textit{Shortest Ping} \cite{Katz-Bassett2006} is a simple delay-based technique. Each target is mapped to the Landmark that is closest to it in terms of the measured RTT.
\textit{GeoPing} as part of IP2Geo \cite{Padmanabhan2001} uses network delay measurements from geographically distributed locations to set up latency vectors.
The corresponding location of the latency vector which is most similar to the target is inferred as geographic location. 

In addition it is possible to further divide measurement based techniques in constrained- and topology-based as well as hybrid approaches. Since hybrid approaches might be capable of integrating also IP mapping based strategies, they are still basically relying on active probing and thus can be considered as measurements based.
%
\textit{Constraint-Based Geolocation (CBG)} \cite{Gueye2006} infers the geographic location of Internet hosts by using multilateration with distance constraints. Hence establishing a continuous space of answers instead of a discrete one.
\textit{Topology-Based Geolocation (TBG)} \cite{Katz-Bassett2006} introduces topology measurements to simultaneously geolocate intermediate routers. Nevertheless TBG is only an enhanced version of the original CBG.
%
%
\textit{Octant} \cite{Wong2007} is a framework for IP Geolocation and the current ``State-of-the-Art'' in terms of active measurements based approaches \cite{Eriksson2011}.
%
Other examples for hybrid or measurement based approaches are HawkEyes \cite{Dahnert2011}, Spotter \cite{Laki2011} and Posit \cite{Eriksson2011}.

Almost all current measurement approaches, in particular the obtained accuracy, are highly depending on Landmarks. The dilemma of using as much as necessary but as few as possible and well distributed Landmarks to reach a highly accurate location estimation, can be considered as \textit{Landmark Problem} \cite{Ziviani2005503,Eriksson2011}.
All notably measurement based approaches use Landmarks for their analyses, but do not provide comprehensive information about optimal selection, positioning or give no attention on this problem at all. In addition, they are using euclidean distances. The work from Ziviani et. al \cite{Ziviani2005503} provides an algorithm for placement of Landmarks. Their proposed linear programming (LP) model is used as a reference for the Landmark location selection. Nevertheless, the presented approach is not suitable for realistic and large scale scenarios. 
Thus, our focus is to determine a predefined amount of Landmarks in a given infrastructure to improve the measurement based IP localization.
\section{Concept}\label{sec:concept}
Our proposed Geolocation service uses a predefined amount of known Landmarks to actively probe the target host. The RTT and hop count from all Landmarks to the target are measured for geographical distance correlation and further calculations. Through multilateration by using these results, the geographical location is inferred.
The underlying problem is the selection and placement of Landmarks out of a given set of possible locations for probing. The Vantage Points have to be identified in the topology with respect to minimize the maximum distance to the surrounding network topology. With better placement of Landmarks, these are closer to the target. As nearer they are, the distance is lower and the variance of measurement results is reduced.


\subsection{Dragoon: Finding Landmarks}\label{sec:landmarks}\label{sec:algorithm}
We introduce a new algorithm Dragoon (Diversification Rectifies Advanced Greedy Overdetermined Optimization N-Dimensions) to find optimized locations for Landmarks, which represent central nodes in a given network topology.
The first placed Landmark usually covers a high amount of nodes, which shows the serious influence of the first placement decision. Nevertheless, an even distribution of Landmarks would be desirable to measure from different directions using several network paths. 
As the target location is not known beforehand, a distributed network of Landmarks is highly recommended. 

Our algorithm starts with a novel initialization process. In the preliminary stage an orientation mark is placed at the optimal position in respect to the distances of the given network topology. Afterwards, the predefined amount of Landmarks is placed using the 2-Approx strategy \cite{Gonzalez85}. 2-Approx calculates for every network node the distance to all placed Vantage Points. It chooses the node with the largest distance to their closest Landmark as the new location to place the next one. 
After the initialization, the algorithm starts with the iterative refinement to find the list of final locations for Landmarks.
%
The algorithm checks all possible locations around the observed Landmark position. The algorithm tests all connected nodes with a direct edge to the current position of the Vantage Point. If the new location improves the overall situation, the algorithm accepts it and replaces the current observed Landmark with the new position. This is done with respect to the specified optimization criterion. In our case, it is the maximum distance counted by hops. If this value is unchanged, the algorithm will use another additional criterion. We use an average or mean criterion to choose between two solutions and to identify an improvement.
In every iteration step, the network nodes are (re)assigned to their closest Landmark. Afterwards, an updated location is calculated for every Vantage Point. This is done with respect to the entire scenario. 
%
In each iteration, every Landmark is allowed to shift its position only once. 
This iterative optimization is repeated until all locations of the Vantage Points do not change any more. 
%
%
%
In comparison to other Landmarks selections, we achieve a better distribution of measurement points with a shorter estimated distance to the target.
\begin{figure}[hbtp]
{
\centering
\includegraphics[width=0.48 \textwidth]{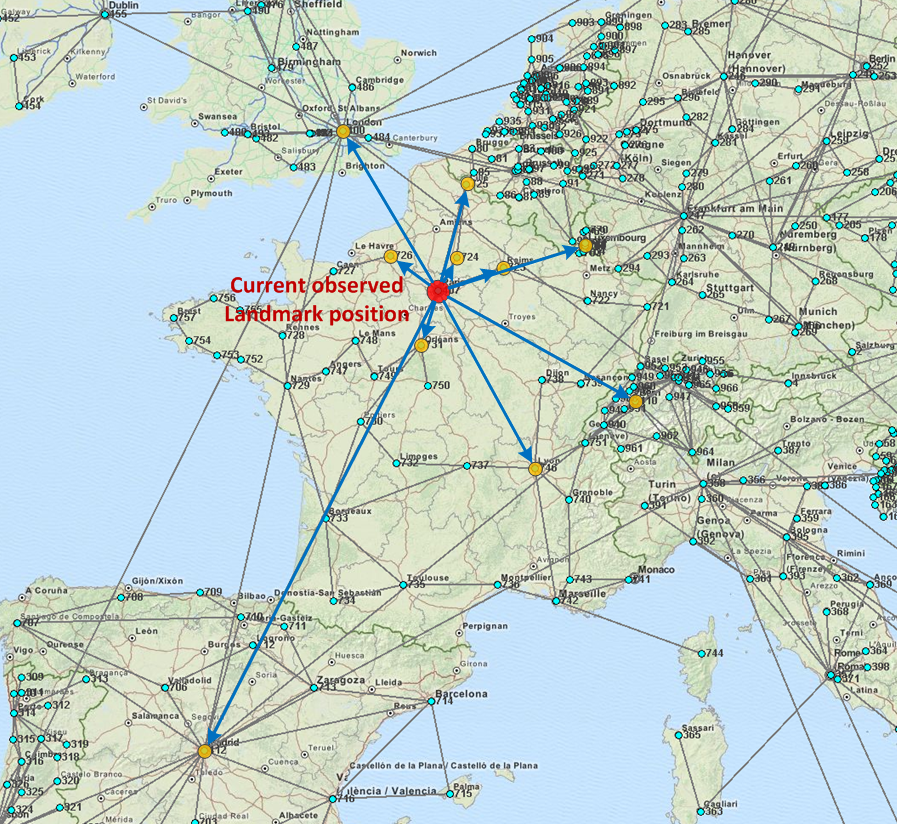}
\caption{Improvement of the current Landmark position (red) with the possibilities (yellow) tried by Dragoon and considered other network nodes (blue).}
\label{fig:DragoonImprovement}
\vspace*{-0.6cm}
}
\end{figure}

%
%
%

\subsection{Conversion of RTT and hop count to distance}\label{conversionRTT}

As we know from the CAIDA data \cite{CAIDA20152} and Section \ref{sec:sota}, the correlation between latency and real distance follows approximately a logarithmic curve. Equation \ref{logarithmicCurve} presents the formula of such a curve fully parametrized, whereas $latency$ is $\frac{RTT}{ 2 }$ subtracted by the $average$ $delay$ of the $detected$ $hops$, which is about $0.1$ $ms$ per hop \cite{Papagiannaki2006}. The reason for the logarithmic correlation is the relatively large transmission delay through the processing units compared to the signal propagation speed in the conductor. This influence is particularly strong in the so called ``Last Mile'', the connection to the end user \cite{Ziviani2005503}. In comparison, most current research work abstract this correlation as a linear function. Such modelling do not take the different Tier network levels into account. In this case it can be considered to be imprecise.
\begin{equation}
distance = p * log_{ e }\left( q * latency + n \right) + m
\label{logarithmicCurve}
\end{equation}

The parameters p, q, n and m  for such a curve are not known and should be calculated for every Landmark individually, because of their unique location in the network topology. For the curve reconstruction and evaluation, we estimate the function based on multiple inter Landmark measurements and curve fitting with a minimized sum of squared-error. 

The distance calculation in our model is based on the WGS84 reference ellipsoid as well as orthodromic distances, also known as great-circle. This is also used by the Global Positioning System (GPS). Thereby, we achieve a much higher accuracy than modelling the earth as a ball without mountains and valleys, rotation flattening effect and applied euclidean distances like in \cite{Dabek2004}. An orthodromic distance corresponds to the shortest distance between two locations on the surface of a sphere, whereas the euclidean space represents the length of a straight line between these locations. Indeed, the path on the surface especially along the network infrastructure is obvious longer than a straight line, which leads to larger distances and wrong estimated target locations. 



\subsection{Lateration}
A geographical location of a target can be estimated using lateration. It uses two known geographical Landmark locations and the estimated distance from each to the target. To increase the precision, multiple Landmarks are probing the targets IP address. As we dealing with distances by the calculation of lateration, the measured RTT and the hop count are converted to a distance using the determined logarithmic function and the principle described in Section \ref{conversionRTT}.
The calculated distance represents the radius $r_{t}$ of a circle or ellipse with the location of Landmarks in the center ($x_{t}$,$y_{t}$):

\begin{equation}
\left( x - x_{ t } \right)^{ 2 } + \left( y - y_{ t } \right)^{ 2 } = r_{ t }
\end{equation}

For the lateration, we calculate the intersection of two circles. This is done without using angles. This is mandatory for a precise modelling, due to the shape of the Earth. 

\begin{figure}[hbt]
\begin{minipage} [hbt]{4.192cm}
\begin{center}
\includegraphics[width=1.13 \textwidth]{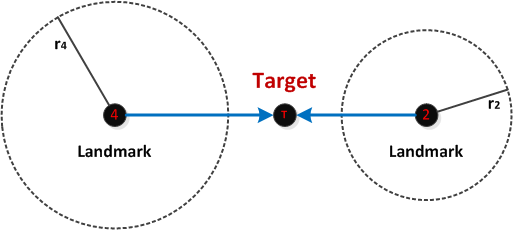}\\
\caption{Zero intersections. Non-overlapping circles.}
\label{fig:circleIntersection0}
\end{center}
\end{minipage}
\hfill
\begin{minipage} [hbt]{4.192cm}
\begin{center}
\includegraphics[width=0.51 \textwidth]{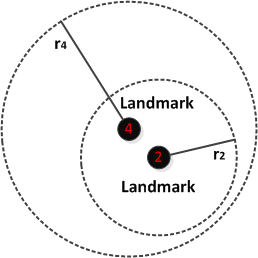}\\
\caption{Zero intersections. Overlapping circles.}
\label{fig:circleIntersection1}
\end{center}
\end{minipage}
	\vspace*{-0.2cm}
\end{figure}
\begin{figure}[hbt]
\begin{minipage} [hbt]{4.192cm}
\begin{center}
\includegraphics[width=0.92 \textwidth]{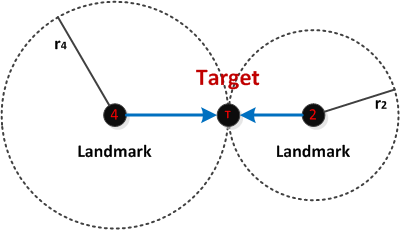}\\
\caption{One intersection point.}
\label{fig:circleIntersection2}
\end{center}
\end{minipage}
\hfill
\begin{minipage} [hbt]{4.192cm}
\begin{center}
\includegraphics[width=0.75 \textwidth]{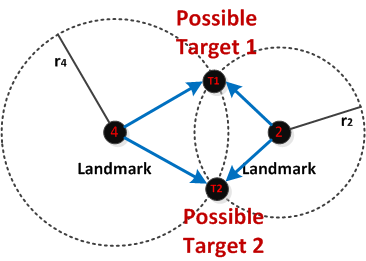}\\
\caption{Two intersection points.}
\label{fig:circleIntersection3}
\end{center}
\end{minipage}
	\vspace*{-0.2cm}
\end{figure}

Two circles can have zero, one or two intersections. In the case of zero intersection point and non overlapping circles, the target location can be estimated in the middle of the space between two Landmarks. If the circles are completely overlapping, it is likely that the target is in the range of the circle with the smaller radius. Otherwise, results from zero intersection point can be disregarded in further calculation. Alternatively, we reduce the radius of the larger circle until we get an intersection point. As we deal with probabilities, the estimated location is more likely to be at this point. In the case of just one intersection point we assume this point as the location of the target. The only inconclusive case is about two intersection points. Here we need further information calculated with support of other Landmarks to decide between the two possible solutions to get the right one. 
The different cases are visualized in the Figures \ref{fig:circleIntersection0} to \ref{fig:circleIntersection3}.

\subsection{Target location estimation}
\begin{figure}[hbt]
\begin{minipage} [hbt]{2.292cm}
\begin{center}
\includegraphics[width=0.95 \textwidth]{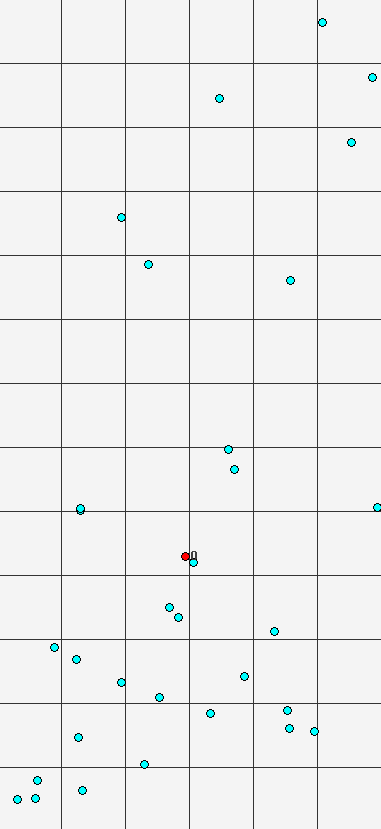}\\
\caption{Intersection point cloud.}
\label{fig:cloud}
\end{center}
\end{minipage}
\hfill
\begin{minipage} [hbt]{6.092cm}
\begin{center}
\includegraphics[width=1.0 \textwidth]{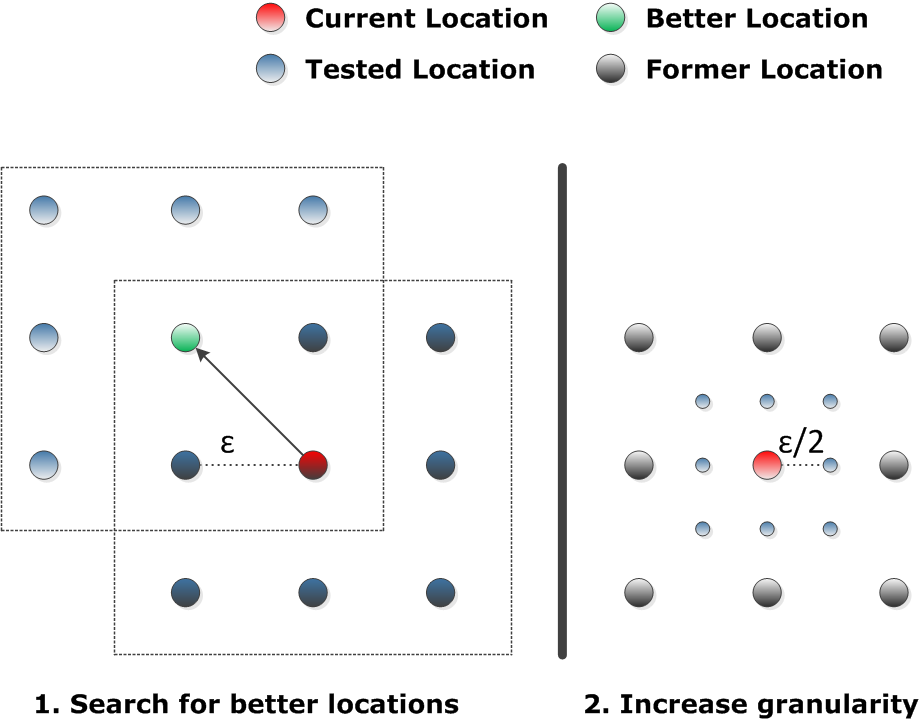}\\
\caption{Iterative optimization stage of the algorithm Dragoon by free placement constraint.}
\label{fig:DragoonImprovement2}
\end{center}
\end{minipage}
	\vspace*{-0.4cm}
\end{figure}

As result of the conducted multilateration, we get a cloud of multiple locations, where the target location is estimated as shown in Figure \ref{fig:cloud}. With our Dragoon algorithm, adapted to a constrained free center placement, we calculate the center location of all points according to the optimization criterion \textit{minimized average distance}. For the free placement constraint, our algorithm tests all points on a grid with a defined distance ($\epsilon$). If one of the tested locations results in a better performance for the overall scenario, it will be accepted. This location is used for the next iteration step. If no location leads to an improvement we successively decrease the granularity of the grid \mbox{($\epsilon_{new}$ := $\frac{\epsilon_{old}}{2}$)}. This process is repeated until the grid distance $\epsilon$ is smaller than the maximal accepted deviation. 
The processing steps of the iterative optimization are shown in Figure \ref{fig:DragoonImprovement2}. The left side illustrates the movement to an improved spot. The right side shows the increased granularity of the grid by bisection. For an improved target location estimation, we filter single points, which are too far away from the other points. Therefor, we iteratively repeat the placement of the center node and filter the points with largest distances from the optimized center location.

\subsection{Measurement}
From a measurement point of view, the end-to-end delay over a fixed path can be split into two components: A deterministic and a stochastic delay \cite{Bovy2002}. The deterministic delay is composed by the minimum processing time at each router, the transmission delay, queuing delay, and the propagation delay. This deterministic delay is fixed for any given path and is taken into account in our concept described in Section \ref{conversionRTT}. The stochastic delay composes the queuing delay at the intermediate routers and the variable processing time as well as buffering at each router that exceeds the minimum processing time. To counter this stochastic delay, several measurements are necessary to get a value close to the theoretical minimal RTT. 

Considering the multiple results, we are only interested in the minimal RTT to get the correct distance and to avoid misleading measurement values caused by circuitous routing. 
Apart from the delay measurements, the hop count on the path has to be determined by tracing the target. This value is used by time measurement and in the calculation in Section \ref{conversionRTT}.

\section{Evaluation}\label{sec:evaluation}
In order to verify the improvements of our introduced concept and algorithm, in terms of selection and positioning of Landmarks, we set up an experiment for geolocating IP addresses.
	\vspace*{-0.2cm}
\subsection{Evaluation Environment and Procedure}
To get a first impression how our algorithm is performing we are focusing on Europe using public available information as well as research data on the Tier 1 topology like \cite{Knight2011}. 
Since the introduced algorithm calculates the optimal position for Landmarks in respect to a given topology, we have to build a set of distributed reference hosts to which we have access to. This is mandatory for probing and determining the geographic location of our target systems. For this purpose we are using the RIPE Atlas Project \cite{Ripeatlas} providing us with over 8200 well-known nodes for probing. 
To avoid confusion the calculated Landmarks are in the following referenced as \textit{Center Nodes}, wheres the reference hosts which we are in the end using for probing are the actual Landmarks. For a first evaluation we use ten Landmarks to cover entire Europe.

The first step is to calculate optimal positions, determined by latitude and longitude, in respect to the given topology. Afterwards we compare these Center Nodes to our set of reference hosts to find direct matches according to latitude and longitude. If no direct match is possible we chose the reference host as actual Landmark, which is closest to the position of the calculated Center Node.
The next step is to measure the RTT and hop count between all Landmarks identified in the previous step. To determine the hop count and the RTT we use ``Paris Traceroute'' and ICMP echo request provided by the RIPE Atlas measurement interface. By using the hop count and the measured minimum delay out of ten measurements, a logarithmic curve is calculated in order to represent a correlation between measured latency and geographic distance. The curve reconstruction is calculated parameter pairwise iteratively with the tool R and the curve fitting method $nls$.

After probing the target IP address from each Landmark, the curve is used to convert the RTT and hop count to a geographic distance. Using the calculated distance and the knowledge of the longitude as well as the latitude of each probing Landmark, Dragoon is able to infer the actual location of the target.

	\vspace*{-0.2cm}
\subsection{Results and Findings}
The Table \ref{tab:AlgoCompare} shows an excerpt from the comparison of estimated target locations obtained by different applied Landmarks, which are identified by Dragoon and 2-Approx.
Since the used scenario is too large for common LP solver, we used the alternative algorithm 2-Approx to the LP in Section \ref{sec:sota}. It illustrates the impact of the selected Landmarks to the results of IP Geolocation.

Nevertheless, in comparison to the solution presented in \cite{Laki2011, Eriksson2011, Laki2009} we achieved better results based on active measurements.
Considering the stochastic delay, the measurements have been conducted between afternoon and early evening. During this time the network load and variance is higher in comparison to other day times. Nevertheless, our applied modelling shows stable results.

\begin{table}[hbtp]
	\centering
	\caption{Comparison of the derivation between the location estimation to the real geographic location using different Landmarks.} 
		\label{tab:AlgoCompare}
	\begin{adjustbox}{max width=\textwidth}
		\begin{tabular}{clll}
			\hline
			\textsc{Target} & \textsc{Dragoon} & \textsc{Reference Algorithm}\\
			\hline
			1 & 117 km & 350 km \\
			2 & 536 km & 1600 km \\
			3 & 108 km & 113 km\\
			\hline
		\end{tabular}
	\end{adjustbox}
	\vspace*{-0.3cm}
\end{table}

\section{Conclusion}\label{sec:conclusion}
In this paper we propose a novel strategy Dragoon to optimize the selection and positioning of Landmarks in a given network infrastructure. Our strategy outperforms existing IP Geolocation approaches based on active measurements in real-world environments. Considering the selection and positing of Landmarks, our algorithm achieve results close to the global optimum. We show the general usability of time measurements for IP localization with high precision based on the selection and position of Landmarks. Reasonable Landmark positions are important for an accurate IP localization service. The closer a Landmark is to a target, the lower are the interferences during the measurements.


\vspace*{-0.3cm}
\section*{Acknowledgment}
We want to thank Sebastian Seeber and  Frank Tietze for providing the needed credits for applying the measurements using RIPE Atlas and many many helpful discussions.
Additional thanks for supporting us goes to RUAG Schweiz AG, Division RUAG Defence - in particular the business unit NEO Services.


\renewcommand\refname{References}
\bibliographystyle{IEEEtran}
\bibliography{IEEEabrv,literature}
\end{document}